  \documentclass[preprint, sort&compress, 3p, 12]{elsarticle}
  \usepackage{graphics,graphicx}
  \usepackage{epsfig}
  \usepackage{amstext}
  \usepackage{amssymb}
  \usepackage{amsfonts}
  \usepackage[colorlinks=true, urlcolor=blue, citecolor=blue, linkcolor=blue, hyperindex]{hyperref}
\hypersetup
{
  pdfauthor   = {Leonardo José Lessa Cirto},
  pdftitle    = {Influence of the interaction range on the thermostatistics of a classical many-body system},
}
   \graphicspath{ {./FIG/} }
 \usepackage{caption}
\newcommand{\rd}{\mathrm{d}}
\newcommand{\de}[1]{\left(#1\right)}

\newcommand{\comub}[1]{\bigl[#1\bigr]}
\newcommand{\Ntil}{ \widetilde{N} }
\newcommand{\media}[1]{\left\langle #1 \right\rangle}
\journal{Physica A: Statistical Mechanics and its Applications}
\begin{document}
\begin{frontmatter}
\title{Influence of the interaction range on the thermostatistics of a classical many-body system}
\author[CBPF]{Leonardo J.L. Cirto\corref{label_corr}}
      \cortext[label_corr]{Corresponding author}
      \ead{cirto@cbpf.br}
\author[UEFS]{Vladimir R.V. Assis}
      \ead{vladimir@cbpf.br}
\author[CBPF,Santa_Fe]{Constantino Tsallis}
      \ead{tsallis@cbpf.br}
\address[CBPF]{Centro Brasileiro de Pesquisas Fisicas and National Institute of Science and Technology for Complex Systems, Rua Xavier Sigaud 150, 22290-180 Rio de Janeiro-RJ, Brazil}
\address[UEFS]{Departamento de Fisica, Universidade Estadual de Feira de Santana, 44031-460 Feira de Santana-BA, Brazil}
\address[Santa_Fe]{Santa Fe Institute, 1399 Hyde Park Road, Santa Fe, NM 87501, USA}
\begin{abstract}
We numerically study a one-dimensional system of $N$ classical localized planar rotators coupled through interactions which decay with distance as $1/r^\alpha$ ($\alpha \ge 0$).
The approach is a first principle one (\textit{i.e.}, based on Newton's law), and yields the probability distribution of momenta.
For $\alpha$ large enough and $N\gg1$ we observe, for longstanding states, the Maxwellian distribution, landmark of Boltzmann-Gibbs thermostatistics.
But, for $\alpha$ small or comparable to unity, we observe instead robust fat-tailed distributions that are quite well fitted with $q$-Gaussians.
These distributions extremize, under appropriate simple constraints, the nonadditive entropy $S_q$ upon which nonextensive statistical mechanics is based.
The whole scenario appears to be consistent with nonergodicity and with the thesis of the $q$-generalized Central Limit Theorem.
\end{abstract}
\begin{keyword}
Metastable state      \sep 
Long-range interaction \sep 
Ergodicity breaking    \sep 
Nonextensive statistical mechanics
\end{keyword}
\end{frontmatter}
\section{Introduction}
More than one century ago, in his historical book {\it Elementary Principles in Statistical Mechanics} \cite{Gibbs1902}, J. W. Gibbs remarked that systems involving interactions such as Newtonian gravitation are intractable within the theory proposed by Boltzmann and himself, due to the divergence of the canonical partition function.
This is of course the reason why no standard temperature-dependent thermostatistical quantities (\textit{e.g.}, a specific heat at finite temperatures) can possibly be calculated for the free hydrogen atom, for example.
Indeed, although the quantum approach of the hydrogen atom solves the divergence associated, for classical gravitation, with small distances, the divergence associated with large distances remains the same. More precisely,
unless a box surrounds the atom, an infinite number of excited energy levels accumulate at the ionization value, which yields a divergent canonical partition function at any finite temperature. This and related questions are commented in \cite{Tsallis2009}, for instance.

Here we report a numerical study of the $\alpha$-XY model~\cite{AnteneodoTsallis1998}, a many-body Hamiltonian system with two-body interactions whose range is controlled by a parameter of the model. More precisely, we  assume a potential which decays with distance as $1/r^\alpha$ ($\alpha \ge 0$).
This model recovers, when $\alpha=0$, the Mean Field Hamiltonian (HMF), a fully coupled many-particle system~\cite{AntoniRuffo1995}, and recovers, in the $\alpha\to\infty$ limit, the first-neighbour XY ferromagnet, a model which is well defined within the traditional thermostatistical scenario of short-range interactions.
Systems with long-range interactions have been attracting particular attention of the  statistical-mechanical community in the last two decades.
This renewed and increasing interest was launched by Antoni and Ruffo in 1995~\cite{AntoniRuffo1995} with their discussion of the HMF model, which  in many aspects mimics traditional long-range systems while bypassing some of its difficulties.

The model focused on here was introduced some years later in~\cite{AnteneodoTsallis1998}.
It is a direct generalization of the HMF by including  a power-law dependence on distance in order to control the range of the interactions\footnote{We refer the reader to the work by Chavanis and Campa~\cite{ChavanisCampa_2010}, where a quite complete list of references about the evolution of this subject can be found.}.
We refer to short-range (long-range) interactions when the potential felt by one rotator of a $d$-dimensional system is integrable (nonintegrable), i.e., when $\alpha/d >1$ ($0 \le \alpha/d \le 1$). A direct consequence of this fact is that the total energy is extensive when $\alpha/d >1$, whereas it is superextensive if $0 \le \alpha/d \le 1$.
As we shall  numerically illustrate, for large values of $\alpha/d$ the system exhibits the standard behaviour expected within Boltzmann-Gibbs (BG) statistical mechanics.
However, we shall also exhibit that when long-range interactions become dominant, i.e., when $\alpha/d<1$, the situation is much more complex.
It was proposed in 1988 a generalization of the BG statistical mechanics based on a different entropic functional~\cite{Tsallis2009, Tsallis1988, TEMUCO}.
Within this approach, the thermodynamical structure (free energy, temperature, etc) can be extended~\cite{Tsallis2009, Andradeetal2010, RibeiroNobreCurado2012, Ribeiro_Nobre_Curado_EPJB_2012, Casas_Nobre_Curado_PRE_2012}.
Some of the numerical results presented in the next sections appear to be in close agreement with 
this theory.
\section{The model}
To transparently extract the deep consequences of Gibbs' remark, in the present paper we focus on the influence of the range of the interactions within an illustrative isolated classical system, namely the $\alpha$-XY model~\cite{AnteneodoTsallis1998}.
This model consists of a $d$-dimensional hypercubic lattice of $N$ interacting planar rotators, whose Hamiltonian is given by
\begin{equation}
{\cal H}=\frac{1}{2}\sum_{i=1}^N p_i^2 + \frac{1}{2}\sum_{i = 1}^N \sum_{{j=1}\atop{j\neq i}}^N \frac{1-\cos(\theta_i - \theta_j)}{r_{ij}^\alpha} \;\;\;(\alpha \ge 0) \,,
\label{eq:Hamiltonian_No_Bar}
\end{equation}
with periodic boundary conditions. Each rotator is characterized by the angle $\theta_i\in[0,2\pi)$ and its canonical conjugate momentum $p_i$. Without loss of generality we have considered unit moment of inertia, and unit first-neighbor coupling constant; $r_{ij}$ measures the (dimensionless) distance between rotators $i$ and $j$, defined as the minimal one given the periodic boundary conditions.

For $d=1$, $r_{ij}$ takes the values 1, 2, 3...; for $d=2$, it takes the values 1, $\sqrt{2}$, 2, \ldots; for $d=3$, it takes the values 1, $\sqrt{2}$, $\sqrt{3}$, 2, \ldots.
Notice that, in contrast with Newtonian gravitation, the potential in Hamiltonian~(\ref{eq:Hamiltonian_No_Bar}) does not diverge at short distances since the minimal distance, in any dimension, is always the unit.
The kinetic term in~(\ref{eq:Hamiltonian_No_Bar}), proportional to $p_i^2$, is the traditional one, but the interaction term is long-range for $\alpha \leq d$, which makes the internal energy per particle to diverge in the thermodynamic limit.
Following \cite{Jund_Kim_Tsallis_PRB_1995}, this property can be seen by realising that the energy {\it per particle} of the interaction term varies with $N$ like~\cite{TamaritAnteneodo2000,CampaGiansantiMoroni2000_PRE}:
\begin{equation}
{\widetilde N} \equiv \frac{1}{N}\sum_{i=1}^N\sum_{{j=1}\atop{j\neq i}}^N \frac{1}{r_{ij}^\alpha}
= \sum_{j=2}^N\frac{1}{r_{1j}^\alpha}
\end{equation}
In the $\alpha\to\infty$ limit, $\Ntil = 2d$. If $\alpha/d < \infty$, the discussion of the above sum can be conveniently replaced by the discussion of
the following integral~\cite{AnteneodoTsallis1998}:
\begin{equation}
d \int_1^{N^{1/d}}\!\!\!\!\!\rd{r} \, \frac{r^{d-1}}{r^{-\alpha}} =  \frac{N^{1-\alpha/d}-1}{1-\alpha/d}\,,
\end{equation}
which behaves, when $N \to\infty$, like $N^{1-\alpha/d}/(1-\alpha/d)$ if $0 \le \alpha/d <1$, like $\ln N$ if $\alpha/d=1$, and like $1/(\alpha/d -1)$ if $\alpha/d>1$.
In other words, the total internal energy is extensive (in the thermodynamical sense) for $\alpha/d>1$, and nonextensive otherwise.
In order to accommodate to a common practice, we can rewrite the Hamiltonian ${\cal H}$ as follows~\cite{AnteneodoTsallis1998}:
\begin{equation}
\bar{\cal H}=\frac{1}{2}\sum_{i=1}^N p_i^2 + \frac{1}{2{\widetilde N}} \sum_{i = 1}^N \sum_{{j=1}\atop{j\neq i}}^N \frac{1-\cos(\theta_i - \theta_j)}{r_{ij}^{\alpha}}
= K + V\,,
\label{hamiltonianbar}
\end{equation}
which can now be considered as ``extensive"  for all values of $\alpha/d$, at the ``price" that a microscopic two-body coupling constant becomes now, through ${\widetilde N}$,  artificially dependent on $N$.
However, as shown in~\cite{AnteneodoTsallis1998},
this corresponds in fact to a rescaling of time (hence of $p_i$)\footnote{If we take into account that the momentum~$p_i$ involves a derivative with respect to time~$t$, $\dot{\theta_i}=\frac{\partial \cal H}{\partial p_i} = p_i$, we verify that ${\cal H}=\Ntil \bar{\cal H}$ provided that $t$ is replaced by $\bar{t}=\sqrt{\Ntil} t$.}.
More precisely, this rewriting takes into account the fact that, for all values of $\alpha/d$, the thermodynamic energies (internal, Helmholtz, Gibbs) grow like $N{\widetilde N}$, the entropy, volume, magnetization~$M$, number of particles, etc, grow like $N$ (\textit{i.e.}, remain extensive for {\it both} regions above and below $\alpha/d=1$), and the temperature~$T$, pressure, external magnetic field, chemical potential, etc, must be scaled with ${\widetilde N}$ in order to have {\it finite} equations of states \cite{Tsallis2009}.
The correctness of this (conjectural) scaling was numerically shown in~\cite{TamaritAnteneodo2000} for the present specific $d=1$ system,
and has been profusely verified in the literature for several other systems, e.g., in ferrofluid~\cite{Jund_Kim_Tsallis_PRB_1995},  fluid \cite{Grigera1996},  magnetic \cite{CannasTamarit1996, SampaioAlbuquerqueFortunato_1997, AndradePinho2005},  diffusive \cite{CondatRangelLamberti2002},  percolation \cite{RegoLucenaSilvaTsallis1999, RegoLucenaSilvaTsallis1999_2} systems, among others (see \cite{Tsallis2009} for an overview).
Also, it was analytically proven~\cite{CampaGiansantiMoroni2000_PRE} (see also~\cite{CampaGiansantiMoroni2003}), for any $d$ and $0<\alpha < d$, that rewriting (\ref{hamiltonianbar}) associates  to the $\alpha$-XY model the same behaviour as that, previously known, of the $\alpha=0$~\cite{AntoniRuffo1995} case.
This universality is clearly exhibited by plotting $T/\Ntil$ and $M$ versus $\media{\cal H}/N\Ntil$ for different values of $\alpha$ and $N$ as in~\cite{TamaritAnteneodo2000}, or, equivalently, $T$ and $M$ versus $\media{\bar{\cal H}}/N$ as in~\cite{CampaGiansantiMoroni2000_PRE, GiansantiMoroniCampa_2002_CSF}.
\subsection{\texorpdfstring{The cases $\alpha = 0$ and $\alpha\to\infty$}{The cases alpha = 0 and alpha--infty}}
Equation~(\ref{hamiltonianbar}) unifies two models that have been frequently studied separately, namely the cases  $\alpha=0$ and $\alpha \to \infty$.
The particular instance $\alpha=0$ recovers the HMF model~\cite{AntoniRuffo1995}.
If we conveniently note the potential $V$ as $V_{\alpha}$, we straightforwardly verify that the HMF potential corresponds to
\begin{equation}
V_{0} =
\frac{1}{2{N}} \sum_{i = 1}^N \sum_{j=1}^N\,\comub{1-\cos\de{\theta_i - \theta_j}} \,.
\end{equation}
The restriction $i\neq j$ is not necessary anymore, and we have approximated  $\Ntil=N-1 \sim N$.
The other particular case corresponds to first-neighbour interactions.
Indeed, if we take the $\alpha \to \infty$ limit in Eq.~(\ref{hamiltonianbar}) and considering periodic boundary conditions we obtain:
\begin{equation}
V_{\infty} =
\frac{1}{2} \sum_{i = 1}^N \,\comub{1-\cos\de{\theta_i - \theta_{i+1}}} \,.
\label{eq:Hamiltonian_First_Neighbour}
\end{equation}
The partition function for the first-neighbourhood potential~(\ref{eq:Hamiltonian_First_Neighbour}) was calculated by Mattis~\cite{Mattis1984} using the transfer matrix technique.
The caloric curve ($T$ vs. $u$) is therefore straightforwardly obtained.
For our purpose here it is sufficient to recall that, for $u=0.69$, $T\approx0.7114$.
This value was correctly recovered in our numerical simulations for $\alpha$ sufficiently large, as exhibited in figure~\ref{fig1} for $\alpha=10.0$.
\section{Short- and long-range regimes: Lyapunov exponents}
At the fundamental state, all rotators are parallel, say $\theta_i=0, \, \forall i$, which corresponds to the ferromagnetically fully ordered case.
At high enough energies, the values of  $\{ \theta_i \}$ are randomly distributed, which corresponds to the paramagnetic phase.
In between, for $\alpha < d$, a second order phase transition occurs at a critical temperature $T_{\textrm{c}}=1/2$ corresponding to a critical energy $u_{\textrm{c}}=3/4$~\cite{AntoniRuffo1995, CampaGiansantiMoroni2000_PRE}; the order parameter is the vector magnetization $\mathbf{M}=1/N\sum_{i=1}^N \mathbf{m}_i$, 
where $\mathbf{m}_i= \de{\cos\theta_i, \sin\theta_i}$.

In addition to the above, it has already been shown that, at the special value $\alpha/d=1$, frontier between  the long- and short-range regimes, the dynamical behaviour sensibly changes. Indeed, for $N \to\infty$ and energies corresponding to the paramagnetic region, the largest Lyapunov exponent of the many-body system remains finite and positive for $\alpha/d >1$, whereas gradually vanishes for $0 \le \alpha/d \le 1$.
It vanishes like $N^{-\kappa}$, where $\kappa(\alpha/d)$ decreases from a positive value (close to 1/3) to zero when $\alpha/d$ increases from zero to 1, and remains zero for $\alpha/d \ge 1$.
It is interesting to emphasize that $\kappa$ does not independently depend on $(\alpha,d)$, but only on the ratio $\alpha/d$ \cite{AnteneodoTsallis1998,CampaGiansantiMoroniTsallis2001,CabralTsallis2002}.
Consistently with the fact that, for all values of the energy per particle $u$ in the paramagnetic region, the Lyapunov exponents vanish in the limit $N\to\infty$, $\kappa$ does not depend on $u$.
\section{Numerical procedure and results}
\label{Sec:Numerical_Procedure}
Let us present now the microcanonical molecular-dynamical results that we have obtained for the $d=1$ Hamiltonian (\ref{hamiltonianbar}) with fixed $(N,u)$, the total energy being $Nu$.
To integrate the $2N$ equations of motion we used the Yoshida $4th$-order symplectic algorithm~\cite{Yoshida_1990,fftw}
with an integration step chosen in such a way that the total energy is conserved within a relative fluctuation smaller than $10^{-5}$.
Some of our present results were also checked through the standard Runge-Kutta scheme.
The class of the initial configurations that we run is the so-called water-bag: all rotators started with the same angle $\theta_i = 0 \,, \forall i$, and each momentum $p_i$ is drawn at random from a uniform distribution.
We rescaled all the $p_i$'s in order to precisely achieve the total desired energy $u$ as well as zero total angular momentum, resulting in a uniform distribution with width $2\sqrt{6u}$ and zero mean.
\begin{figure}[h!]
\begin{center}
  \includegraphics[width=0.94\linewidth]{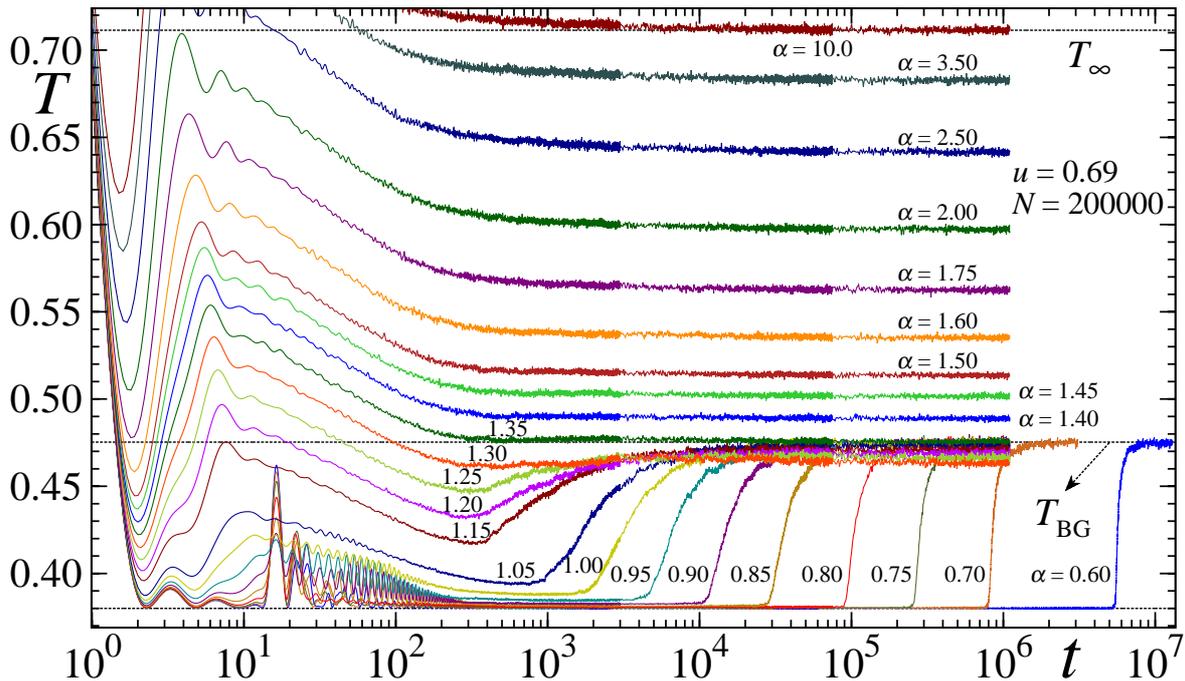}
\end{center}
\caption
{
Time dependence of the kinetic temperature $T(t) \equiv 2K(t)/N$ for a water bag typical {\it single}  initial condition for $(u,N)=(0.69,200000)$ and various values of~\,$\alpha$.
The upper horizontal line, at~\,$T_\infty=0.7114\ldots$, corresponds to the BG thermal equilibrium temperature of the~\,$\alpha \to \infty$ model at~\,$u=0.69$~\cite{Mattis1984}.
The middle (lower) horizontal line, at $T \simeq 0.475$ ($T=0.380$), indicates the BG thermal equilibrium temperature (the QSS base temperature, corresponding to zero magnetization), at $u=0.69$ and $0 \le \alpha  < 1$~\cite{AntoniRuffo1995,CampaGiansantiMoroni2000_PRE}.
In the range $1\leq \alpha <\infty$, no analytical solution is available, as far as we know.
}
\label{fig1}
\end{figure}
\subsection{Temperature and momentum distribution}
We present in figure~\ref{fig1} the instantaneous kinetic ``temperature" $T(t) \equiv 2 K(t)/N$, where $K(t)$ is the time-dependent total kinetic energy of Hamiltonian $\bar{\cal H}$.
As verified many times in the literature, a quasi-stationary state (QSS) exists for $0 \le \alpha/d<1$ and $u \simeq 0.69$, after which a crossover is observed to a state whose temperature coincides with that analytically obtained within Boltzmann-Gibbs (BG) statistical mechanics~\cite{AntoniRuffo1995, GiansantiMoroniCampa_2002_CSF}.
Sufficiently after the QSS period, whose lifetime appears to diverge with increasing~$N$,
the kinetic temperature of the system fluctuates around its BG value as time increases.
The temporal mean value calculated within this stable region is noted $T_{\textrm{kin}}  \equiv  \langle T(t)\rangle = \langle 2K(t)/N\rangle$, and is represented by the full red points in figure~\ref{fig4}.

\begin{figure}\begin{center}
\includegraphics[width=0.85\linewidth]{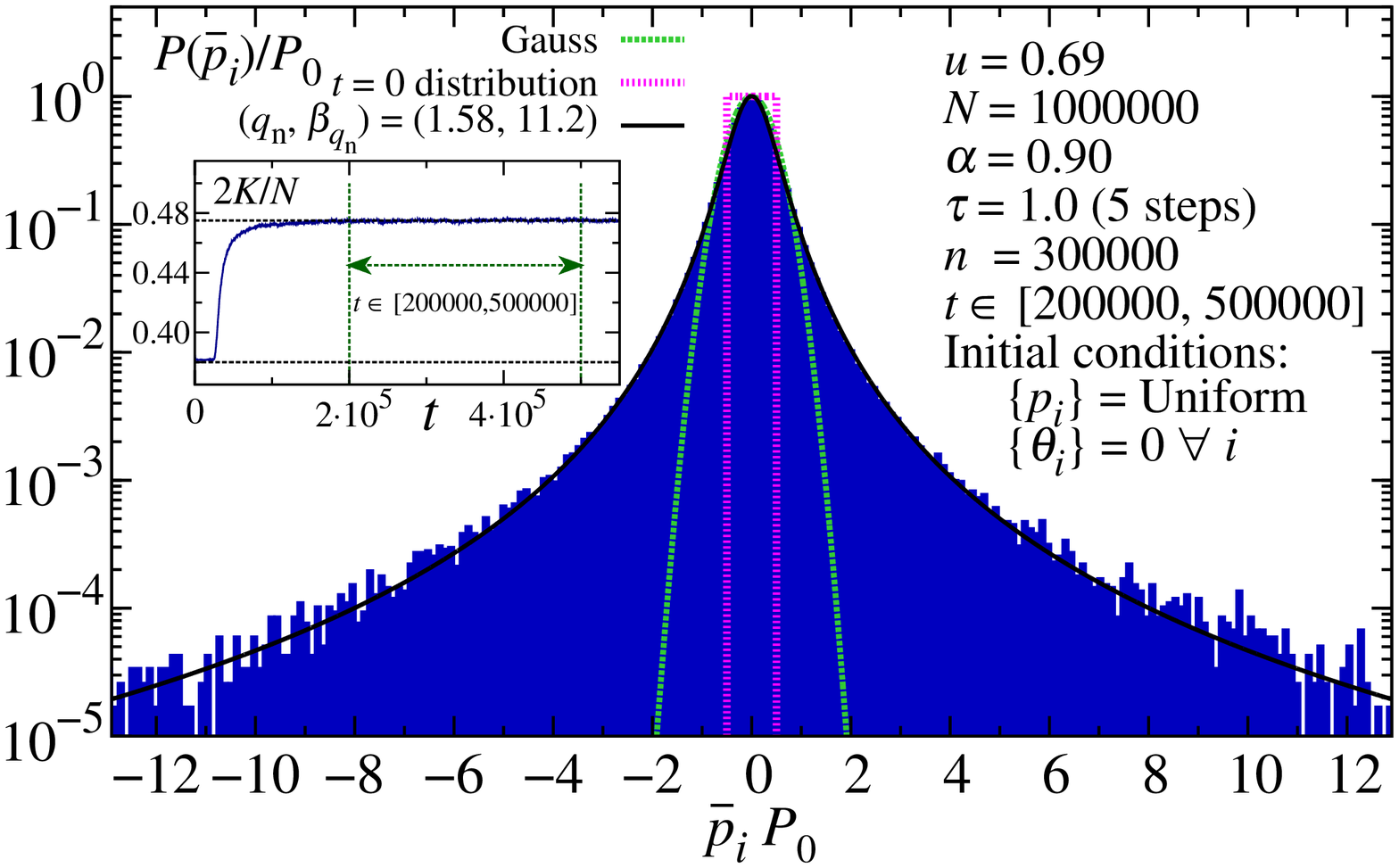} \\
\vspace{0.40cm}
\includegraphics[width=0.85\linewidth]{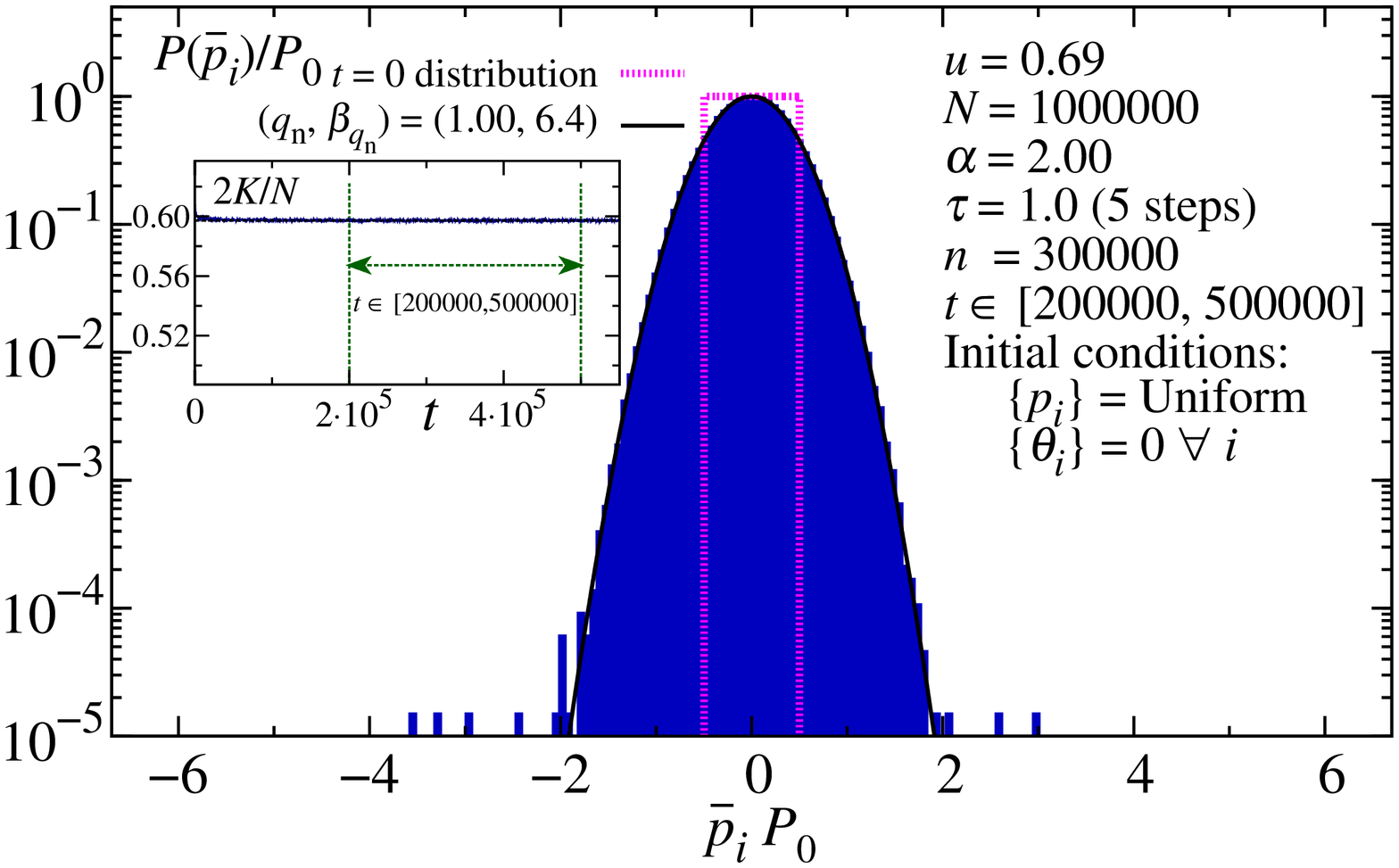}
\end{center}
\caption{A typical single-initial-condition one-momentum distribution $P(p)$ for $N=10^6$, $u=0.69$, $\tau=1$ (corresponding to 5 molecular-dynamical algorithmic steps), calculated in the region $[t_{\textrm{min}},t_{\textrm{max}}]=[200000,500000]$ for $\alpha=0.9$ (top plot), and $\alpha=2.0$  (bottom plot).
The upper temperature indicated in the $\alpha=0.9$ inset coincides with that analytically calculated within BG statistical mechanics, namely $T_{\textrm{kin}} \equiv \langle 2K(t)/N\rangle \simeq 0.475$. The horizontal line of the $\alpha=2.0$ inset corresponds to the time average calculated numerically; indeed, analytical solutions are only available  for $\alpha<1$ \cite{GiansantiMoroniCampa_2002_CSF} and in the $\alpha\to\infty$ limit \cite{Mattis1984}.
The continuous curves correspond to $P(\bar p)/P_0=e_{q_{\textrm{n}}}^{-\beta_{q_{\textrm{n}}}^{(P_0)} [\bar pP_0]^2/2}$ with $(q_{\textrm{n}},\beta_{q_{\textrm{n}}}^{(P_0)})= (1.58,11.2)$ for $\alpha=0.9$ and $(1.0,6.4)$ for $\alpha = 2.0$.
The value of $q_{\textrm{n}}$ for $\alpha=0.9$ corresponds to the red open circle in figure~\ref{fig3}; notice that  $1/\beta_{q_{\textrm{n}}}^{(P_0)} \ne T$ in this case.
Each distribution has been rescaled with its own~$P_0$.}
\label{fig2}\end{figure}

It has been long thought that, after this crossover, the system consistently adopts a BG distribution in Gibbs $\Gamma$ space, and therefore a Maxwellian distribution for~$P(p_i)$.
The facts that we now present reveal a much more complex situation, where robust $q_{\textrm{n}}$-Gaussians (or distributions numerically very close to them) emerge before the crossover (just before for most realizations of the initial conditions, but also quite before for not few of them) and remain so for huge times (as long as our longest runs); $\textrm{n}$~stands for {\it numerical}.
This unexpected phenomenon occurs for $u$ both below and above $u_{\textrm{c}}=3/4$, and for $\alpha$ both below and above $\alpha=1$ (up to $\alpha \simeq 2$).
Let us emphasize that these {$q_{\textrm{n}}$-Gaussians} only develop their full shape if sufficient time has been run in order that the apparently stationary state has been attained.
This time is extremely long for $0<u \ll 3/4$ because the system is then almost integrable (indeed, the Hamiltonian can be straightforwardly checked to become very close to that of $N$ coupled harmonic oscillators, by using $\cos(\theta_i - \theta_j) \sim 1-\frac{1}{2}(\theta_i-\theta_j)^2$), and is also extremely long for $u \gg 3/4$ because once again the system is almost integrable (indeed, the Hamiltonian can be straightforwardly checked to become now very close to $N$ independent localized rotators).
Let us detail now how the single-initial-condition one-momentum distributions $P(p)$ are calculated within large time regions where $T$ is nearly constant:
for each value of $i$, we register its $p_i$ at very many (noted $n$) successive times separated by an interval $\tau$, and then,
following the recipe of the Central Limit Theorem (and of its $q$-generalisation \cite{UmarovTsallisSteinberg2008, Hilhorst2010, JaureguiTsallisCurado2011, PlastinoRoccaMilan2012, VignatPlastino2007, HahnJiangUmarov2010}), we
calculate its arithmetic average~\,$\bar{p}_i$ (thus corresponding to the interval $t \in [t_{\textrm{min}}, t_{\textrm{max}}]$ with $t_{\textrm{max}}-t_{\textrm{min}}=n \tau$).
We then plot the histogram for the $N$ arithmetic averages, as illustrated in figure~\ref{fig2}.
Notice that the CLT recipe is nothing but a time average, which frequently corresponds in fact to real experiments.
 \clearpage
\subsection{\texorpdfstring{$q$-Kurtosis}{q-Kurtosis}}
\begin{figure} \begin{center}
 \includegraphics[width=0.85\linewidth]{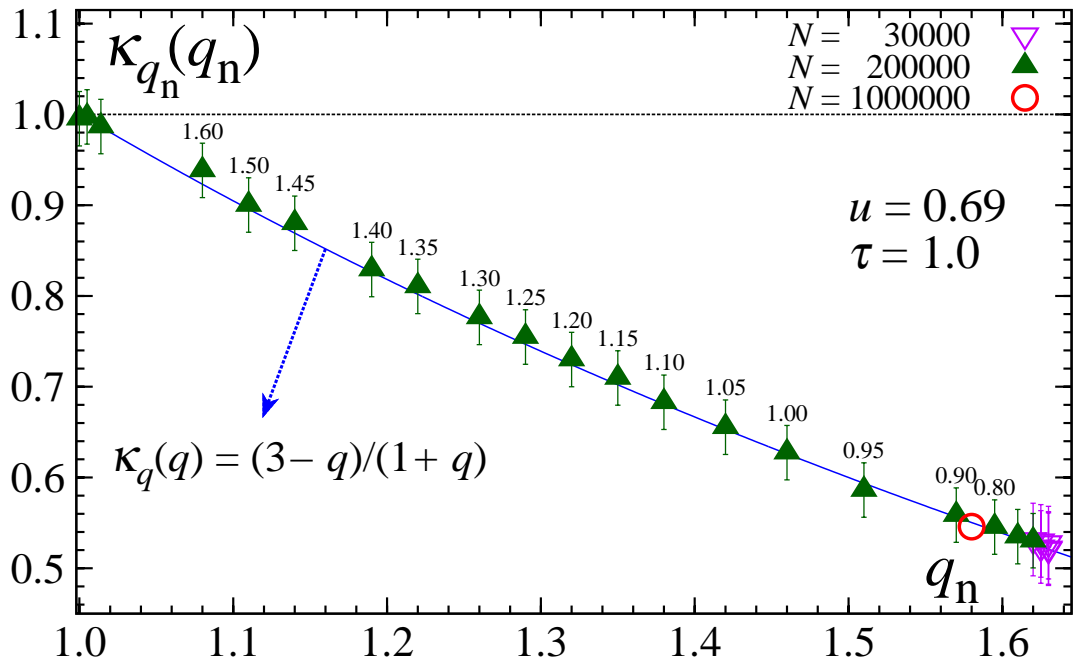}
\end{center}
\caption{$q_{\textrm{n}}$ and $q$-kurtosis $\kappa_{q_{\textrm{n}}}$  that have been obtained from the histograms corresponding to typical values of $\alpha$ (numbers indicated on top of the points).
The red circle corresponds to $\alpha=0.9$ in figure~\ref{fig2}.
The continuous curve $\kappa_q=(3-q)/(1+q)$ is the analytical one obtained with $q$-Gaussians.
Notice that $\kappa_q$ is finite up to $q=3$ (maximal admissible value for a $q$-Gaussian to be normalizable), and that it does not depend on $\beta_{q_{\textrm{n}}}$.  The visible departure from the dotted line at $\kappa_q =1$ corresponding to a Maxwellian distribution, neatly reflects the departure from BG thermostatistics.
}
\label{fig3} \end{figure}
All the histograms that we have obtained for sufficiently large times $t$ are well fitted with $e_{q_{\textrm{n}}}^{-\beta_{q_{\textrm{n}}} p^2/2}$, with $(q_{\textrm{n}},\beta_{q_{\textrm{n}}})$ depending on $(\alpha, u, N, \tau)$ as well as on $(t_{\textrm{min}},t_{\textrm{max}})$ where $e_q^x \equiv [1+(1-q)x]^{1/(1-q)}$  ($q \in {\cal R}; \,e_1^x=e^x$)~\cite{Tsallis2009, Comment_q_Gaussiana}.
To check the quality of the fit we introduce (see  figure~\ref{fig3}) a conveniently $q$-generalized kurtosis (referred to as {\it $q$-kurtosis}), defined as follows (see \cite{TsallisPlastinoAlvarezEstrada2009} and references therein):
\begin{equation}
\kappa_q=\frac{\int_{-\infty}^\infty \rd p \,p^4 [P(p)]^{2q-1}  / \int_{-\infty}^\infty \rd p \, [P(p)]^{2q-1}  }{3 \Bigl[\int_{-\infty}^\infty \rd p \,p^2 [P(p)]^{q}  / \int_{-\infty}^\infty \rd p \, [P(p)]^{q}  \Bigr]^2 } \,,
\label{kurtosis}
\end{equation}
where we have used the escort distributions. Escort distributions were, as far as we know, introduced in 1995 by Beck and Sch\"ogl~\cite{Beck_LIVRO_1995} and play a particular role in nonextensive statistical mechanics~\cite{Tsallis2009, TsallisMendesPlastino_1998}.
The mean values associated with these distributions have the remarkable advantage of being finite up to $q=3$, which is precisely the value below which $q$-Gaussians are normalizable, \textit{i.e.} $\int_{-\infty}^\infty \rd p\,P_0 e_{q}^{- \beta_q p^2/2} =1 \,\,(q<3)$.
The use of the standard kurtosis $\kappa_1 = \langle  p^4\rangle / 3 \langle  p^2\rangle^2$ when the probability distribution is a $q$-Gaussian
has the considerable disadvantage that $\langle  p^2\rangle$ diverges for $q \ge 5/3$, and $\langle  p^4\rangle$ diverges for $q \ge 7/5$.
Hence $\kappa_1$ becomes useless for $q \ge 7/5$, and it happens that some of the distributions that we observe do exhibit $q_{\textrm{n}} \ge 7/5$.
If we use a $q$-Gaussian $P(p)$ within equation~(\ref{kurtosis}), we obtain, through a relatively easy calculation
\begin{equation}
\kappa_q(q)=\frac{3-q}{1+q}\,,
\end{equation}
as also shown in figure~\ref{fig3}.
\clearpage
\subsection{\texorpdfstring{$q$ versus $\alpha$}{q versus alpha}}
\begin{figure} \begin{center}
  \includegraphics[width=0.85\linewidth]{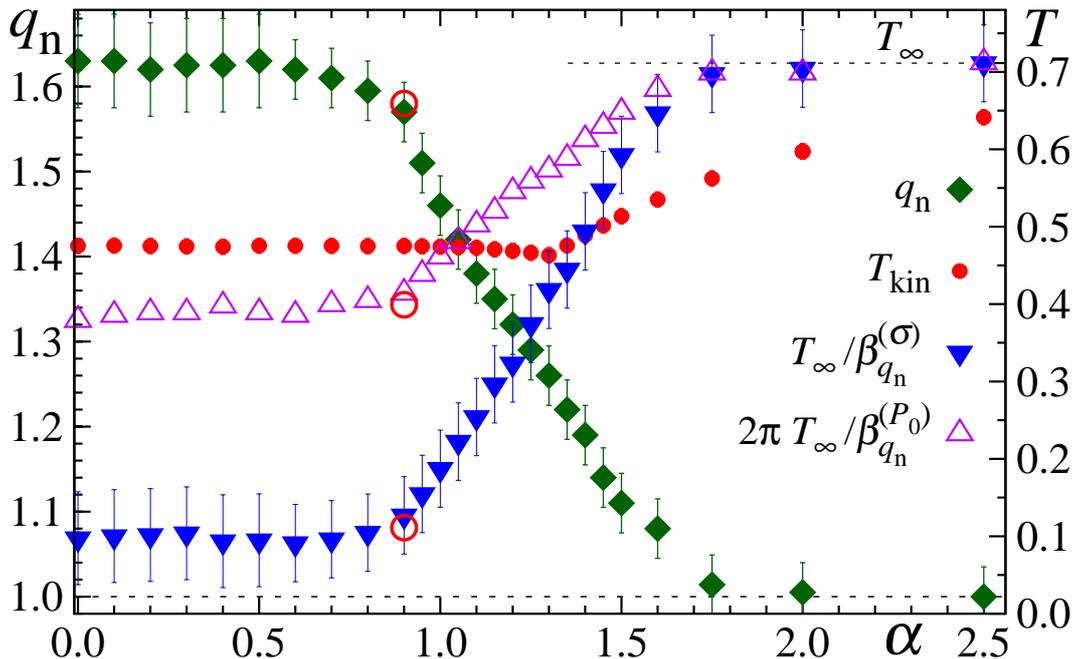}
\end{center}
\caption{$\alpha$-dependences of $(q_{\textrm{n}},\beta_{q_{\textrm{n}}})$ for $(u,\tau,N)=(0.69,1.0,N)$, where $N=200000$ ($N=30000$) for $\alpha \ge 0.6$ ($\alpha \le 0.5$), with $n$ never smaller than 300000.
We have verified the existence of finite-size effects, in particular, for $\alpha$ above and close to unity, $q_{\textrm{n}}$ slowly decreases with increasing $N$.
Notice that $T_{\textrm{kin}}\simeq 0.475$ up to $\alpha \simeq 1.35$, where it starts increasing (red full circles), and, for $\alpha \gg 1$, approaches the analytical value $T_\infty = 0.7114\ldots$\cite{Mattis1984} (by using the values that we have obtained up to $\alpha=40$, we observe that approximatively $ T_\infty-T_\alpha \simeq 0.4 /\alpha^2$ for $\alpha \gg 1$).
The red open circles correspond to the example in figure \ref{fig2} (also indicated in figure \ref{fig3}).
The full (open) triangles have been obtained from rescaled histograms where the momenta have been  divided by the standard deviation $\sigma$ (multiplied by $P_0$, as illustrated in figure~\ref{fig2}); in other words, they both indicate (through two different normalization procedures) a single physical quantity, namely the width of the distribution.
The error bars corresponding to the triangles are of the same order; the error bars of $T_{\textrm{kin}}$ are of the order of the full circles (red).
Naturally, $P_0 \times \sigma$ is nearly constant; to take into account the numerical deviations (from a strict constant) due to parameters such as $(N, n, \tau)$, we have normalized both   $\beta_{q_{\textrm{n}}}^{(\sigma)}$ and $\beta_{q_{\textrm{n}}}^{(P_0)}$  in such a way that the analytical value $T_\infty = 0.7114 \ldots$ is recovered.
}
\label{fig4} \end{figure}
\begin{figure}\begin{center}
\includegraphics[width=0.85\linewidth]{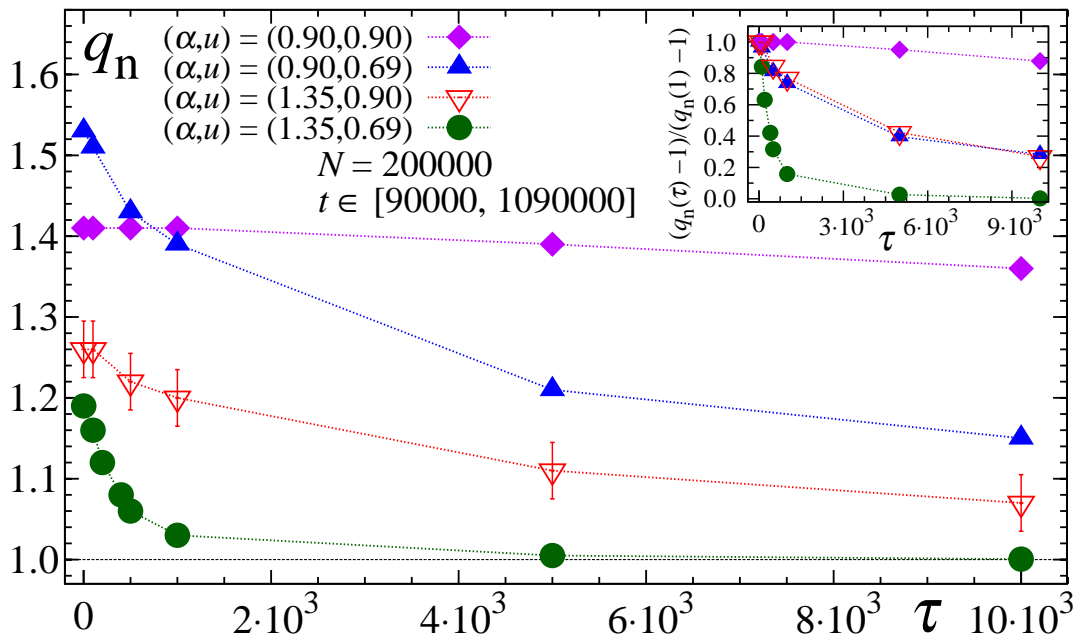}
\end{center}
\caption{$\tau$-dependence of $q_{\textrm{n}}$ for $N=200000$, $[t_{min}, t_{max}]=[90000,1090000]$ (hence $n=1000000$ for $\tau=1$), and typical values of $u$ above and below the critical value $u_{\textrm{c}}=0.75$, and of $\alpha$ above and below the special value $\alpha =1$ (see \cite{AnteneodoTsallis1998}).
All the error bars are of the same order of those indicated on the red empty triangles.
{\it Inset:} $\tau$-dependence of $[q_{\textrm{n}}(\tau)-1]/[q_{\textrm{n}}(1)-1]$.
}
\label{fig5}\end{figure}

In figures~\ref{fig4}  and \ref{fig5} we illustrate $(q_{\textrm{n}},\beta_{q_{\textrm{n}}})$ as functions of $(\alpha, u, \tau)$ for large values of~$N$.
All the $(u,\tau)=(0.69,1)$ results for $q_{\textrm{n}}$
have been also reported in figure \ref{fig3}. One of the interesting features that we can observe is that in all cases $q_{\textrm{n}}$ approaches the BG value $q=1$ when $\tau$ increases.
However, this approach is nearly exponential for $(\alpha <1,\,u<0.75)$,   $(\alpha >1,\,u>0.75)$, and $(\alpha >1,\,u<0.75)$, whereas it is extremely slow for  $(\alpha <1,\,u>0.75)$ (notice that, in the latter case, $q_{\textrm{n}}$ exhibits a zero slope with regard to $\tau$ at $\tau =1$), precisely the region where the largest Lyapunov exponent approaches zero with increasing~$N$
(we remind that the $d=1$ critical point for $0 \le \alpha < 1$ is known to be $u_{\textrm{c}}=3/4$).
This suggests the following nonuniform convergence: $\lim_{N \to\infty} \lim_{\tau \to\infty} q_{\textrm{n}}(\alpha,u,N,\tau)=1$ ($\forall \alpha$), whereas   $\lim_{\tau \to\infty} \lim_{N \to\infty} q_{\textrm{n}}(\alpha,u,N,\tau)>1$ (for $0 \le \alpha <1$).
Lack of computational strength has not allowed us to directly verify this conjecture.
This leaves as an interesting open question whether $\lim_{\tau \to\infty} \lim_{N \to\infty} q_{\textrm{n}}(\alpha,u,N,\tau)$ recovers $\lim_{N\to\infty}q_{\textrm{n}}(\alpha,u,N)$, where the latter would correspond to successive approximations for increasingly large $N$.

For all $\alpha$, our numerical values of $q$ are always larger or equal to unit.
This is not mandatory for Hamiltonian systems, and neither for maps at the edge of chaos.
For example, for the Fermi-Pasta-Ulam finite-size chains, values of $q$ both above and below unity were longstandingly observed in different regions of phase space~\cite{LeoLeoTempesta2010}
 (see also~\cite{BrusselsGroupe_1}, where it is argued -- disputably though -- that $q$-exponentials with values of $q$ above unity are not admissible within nonextensive statistics for classical many-body Hamiltonians).
For the case of maps at their edge of chaos, studies exhibiting values of $q$ both above and below unity are also available~\cite{Tirnakli_et_al_2009, Pluchino_Rapisarda_Tsallis_PRE_2013}.
Finally, $q$-exponential distributions of energy can also be seen in \cite{Campisi_EPL_2012} for $q$ both above and below unity.
\subsection{Angle distribution}
\begin{figure}\begin{center}
\includegraphics[width=0.85\linewidth]{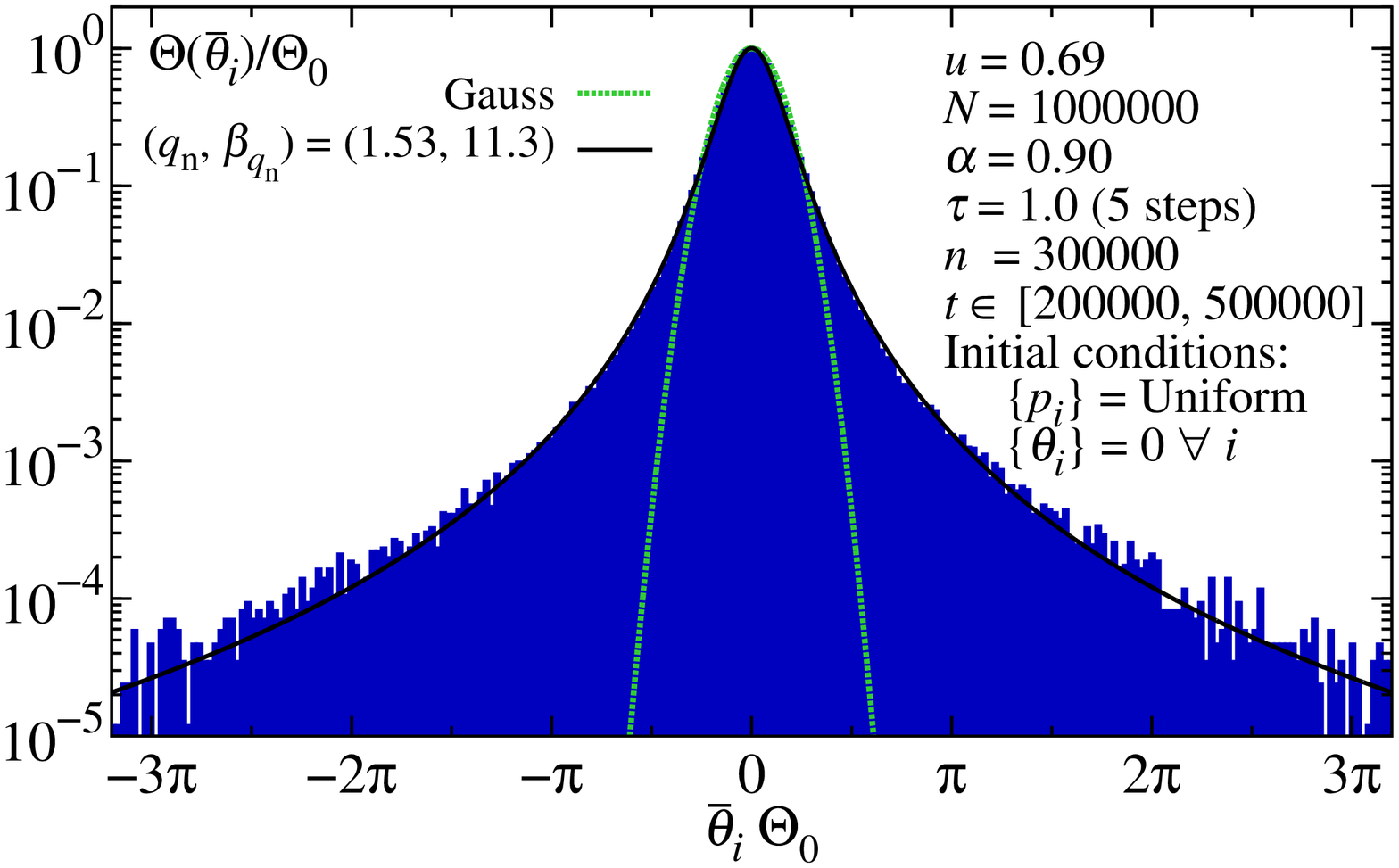} \\
\vspace{0.40cm}
\includegraphics[width=0.85\linewidth]{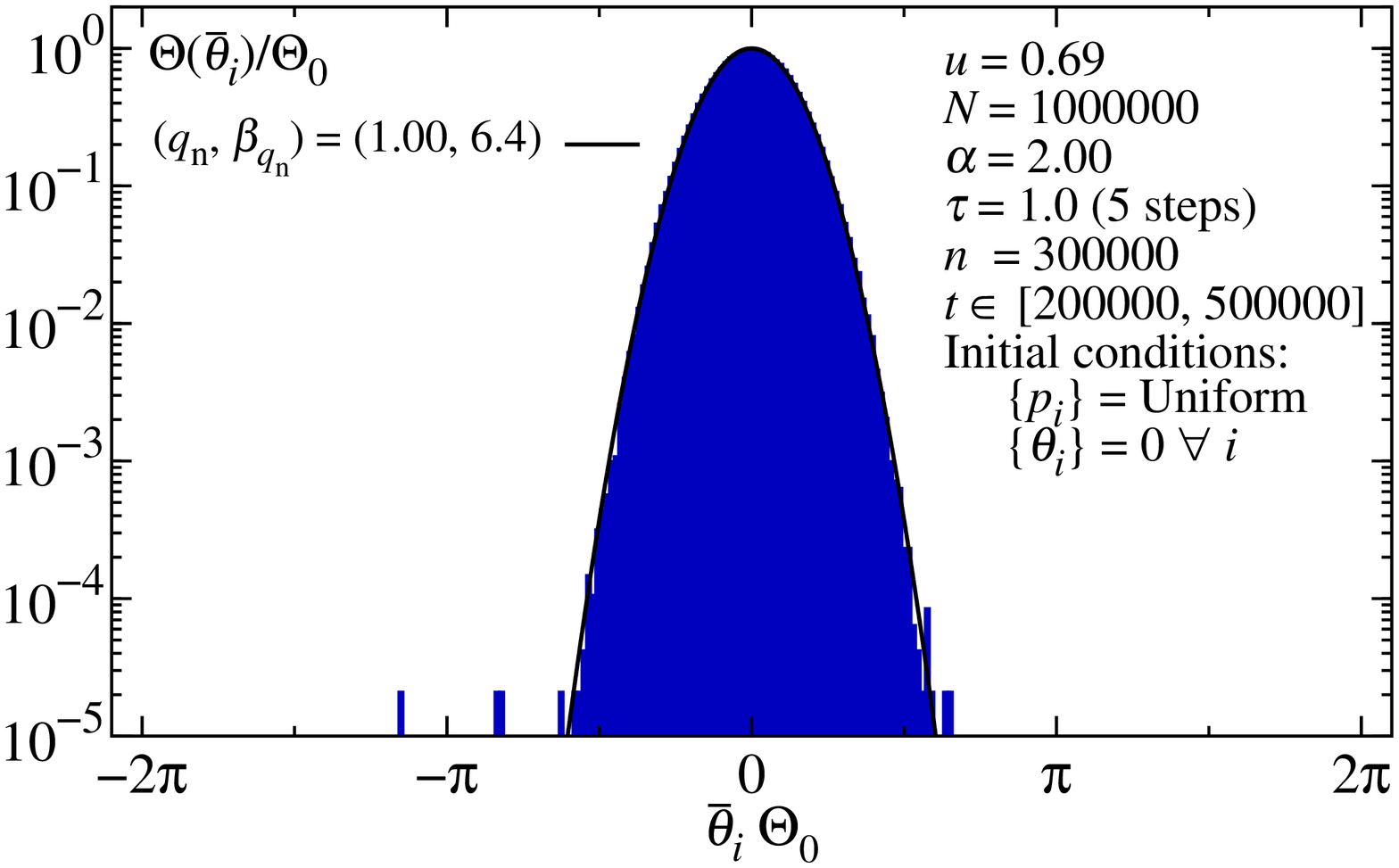}
\end{center}
\caption{A typical single-initial-condition one-angle distribution $\Theta(\theta)$ for exactly the same conditions of figure~\ref{fig2} ($\alpha=0.9$ for the top plot; $\alpha=2.0$ for the bottom plot).
Notice that the fitting  parameters $(q_{\textrm{n}},\beta_{q_{\textrm{n}}})$ of the present $q$-Gaussian $\Theta(\bar \theta)=\Theta_0e_{q_{\textrm{n}}}^{-\beta_{q_{\textrm{n}}} [\bar \theta\Theta_0]^2/2}$ (continuous curve) practically coincide with those of figure~\ref{fig2}.
}
\label{figangle}\end{figure}

To further clarify this new thermostatistical scenario it is helpful to analyse the behaviour of the angles~$\theta_i$'s.
In figure~\ref{figangle} we present the angle distributions obtained in exactly the same conditions of the previously shown momenta distributions (see figure~\ref{fig2}).
The numerical solution of the equations of motion of system~(\ref{hamiltonianbar}) provides an unbound domain
for the canonical variables~\,$\theta_i$'s, namely $\theta_i(t)\in(-\infty,+\infty)$.
Notice though that the dynamics itself depends solely on the  values of the phase modulo $2\pi$ since
angles~$\theta_i$'s appear as arguments of sine and cosine functions.
An unbound representation for the angles is quite convenient since it directly reflects the continuous rotations of the rotors (see also \cite{MoyanoAnteneodo2006}).
\subsection{Time versus Ensemble Averages}
\begin{figure}
\begin{center}
\includegraphics[width=0.99\linewidth]{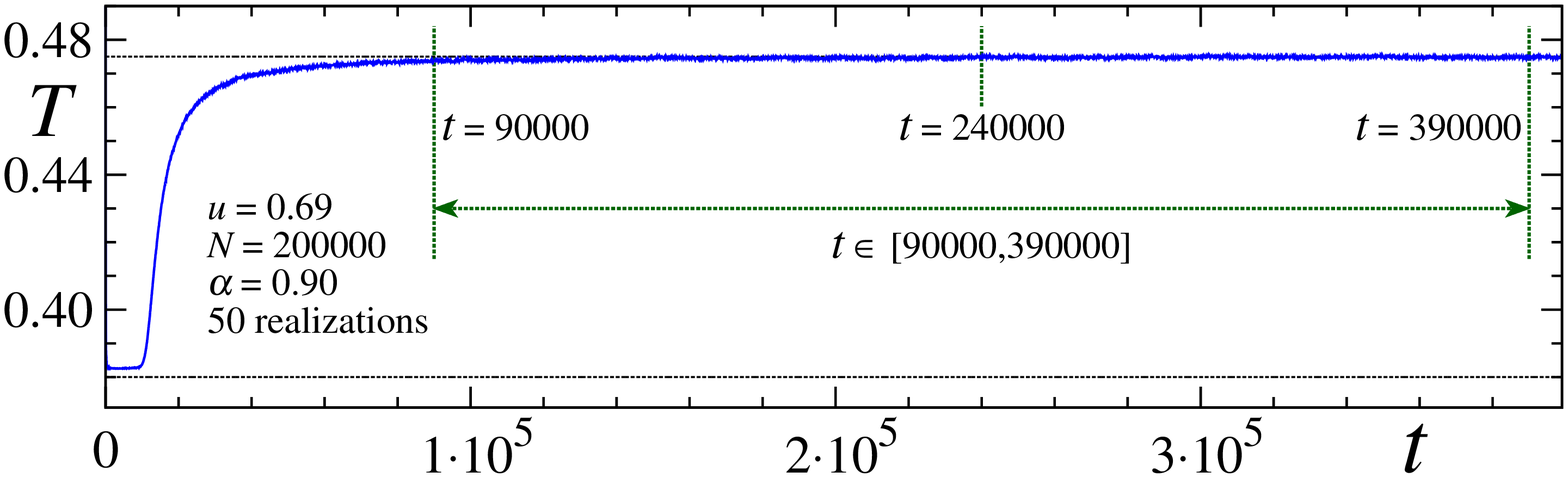}\\
\vspace{0.70cm}
\includegraphics[width=0.49\linewidth]{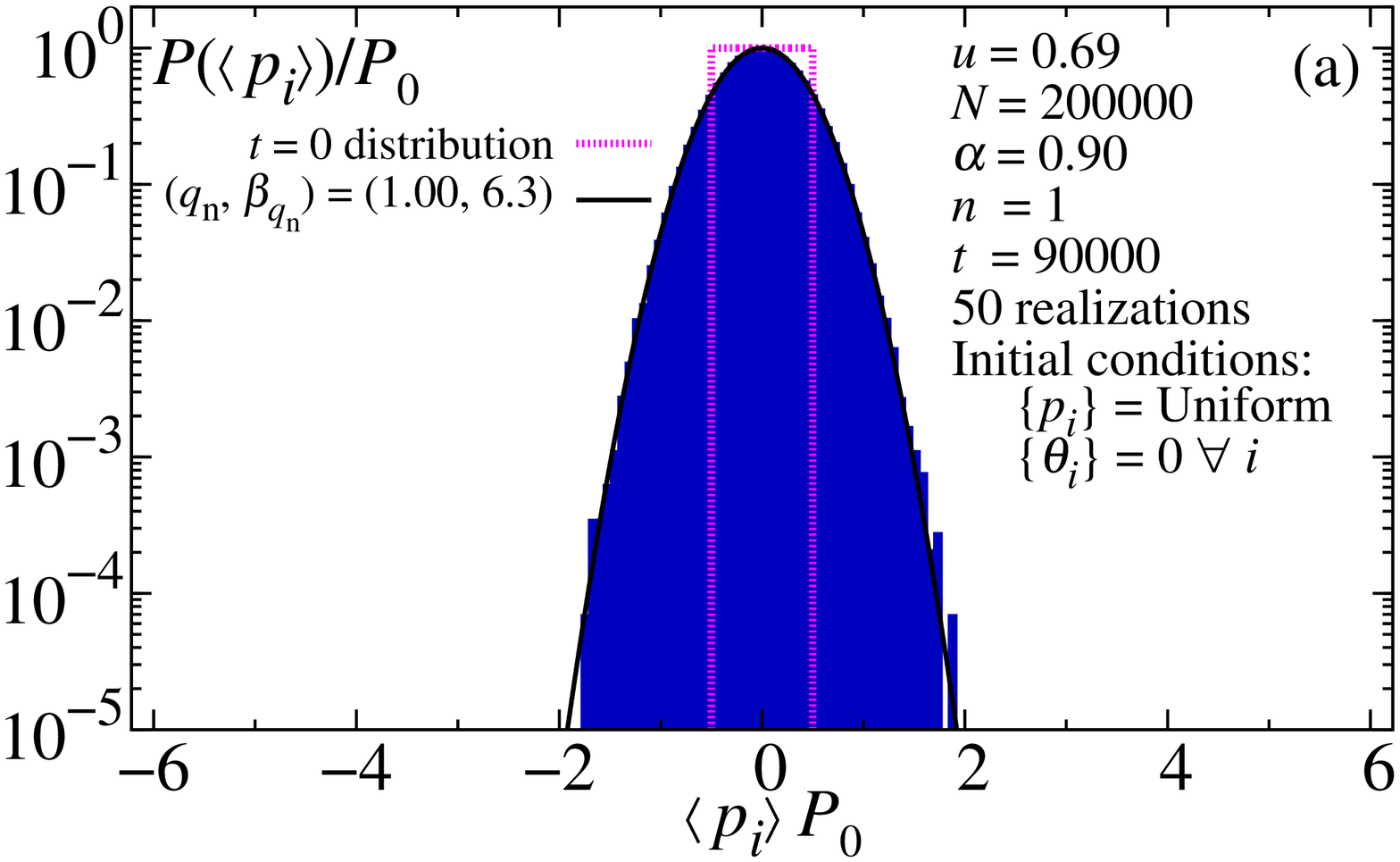}
\includegraphics[width=0.49\linewidth]{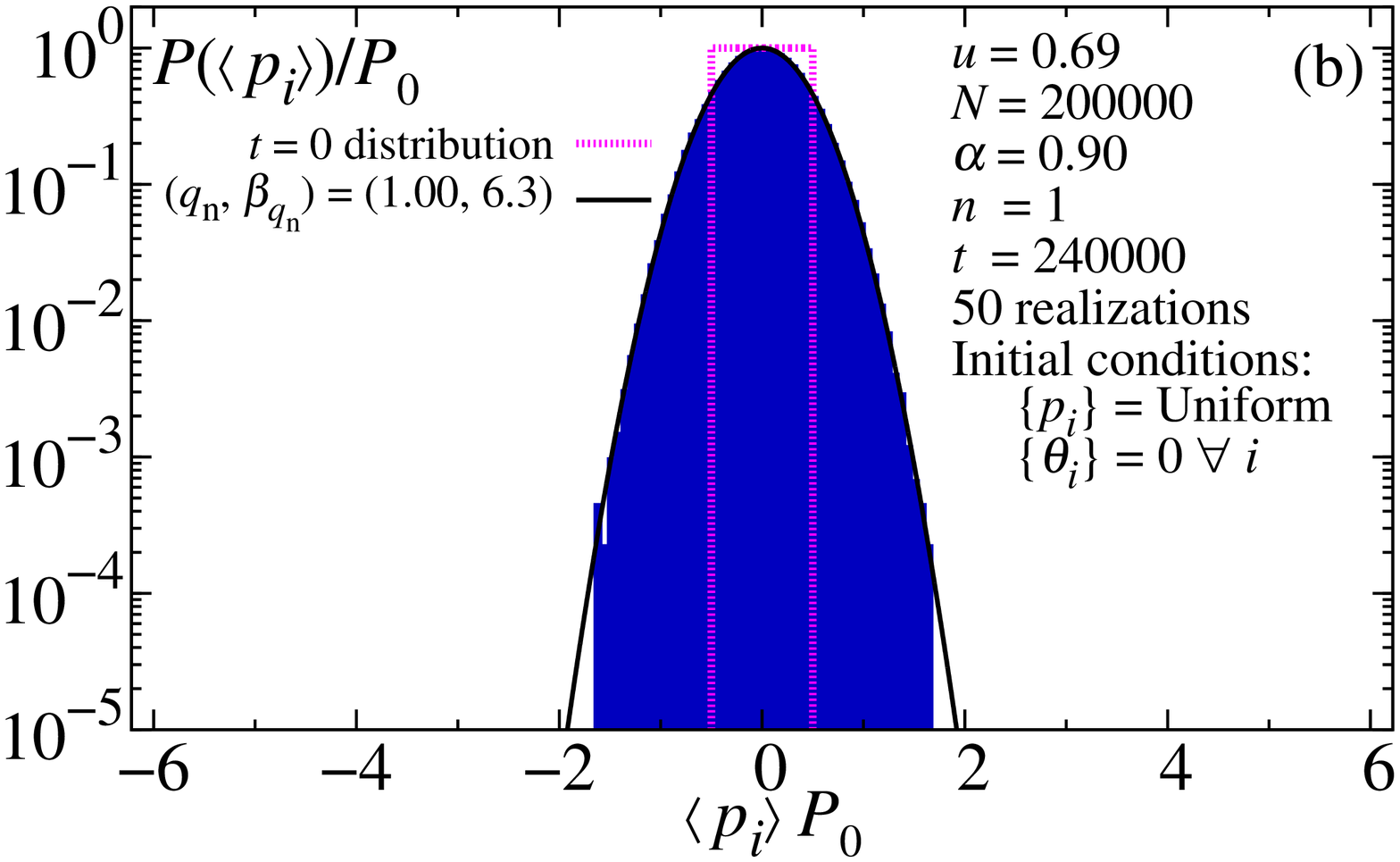}
\includegraphics[width=0.49\linewidth]{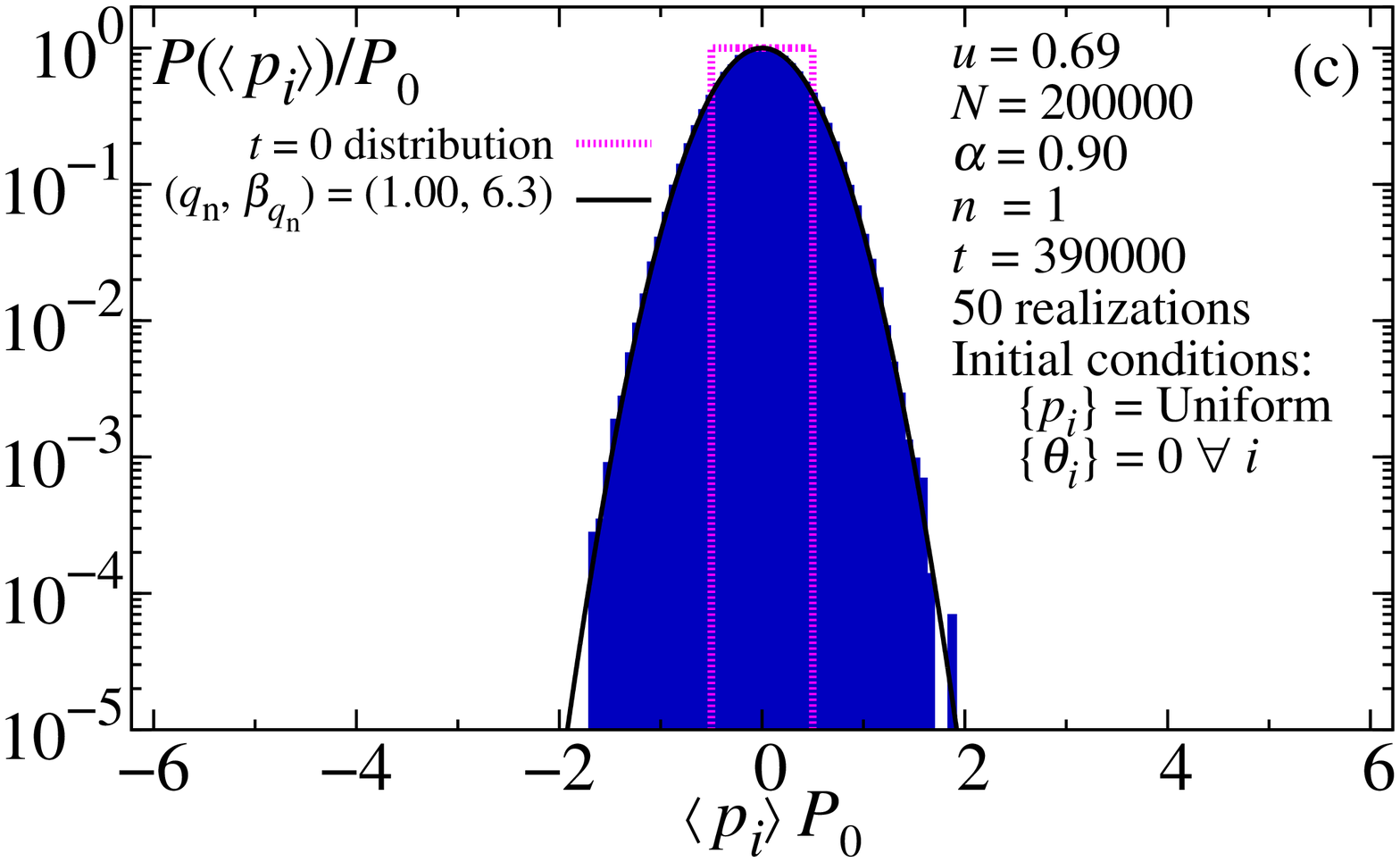}
\includegraphics[width=0.49\linewidth]{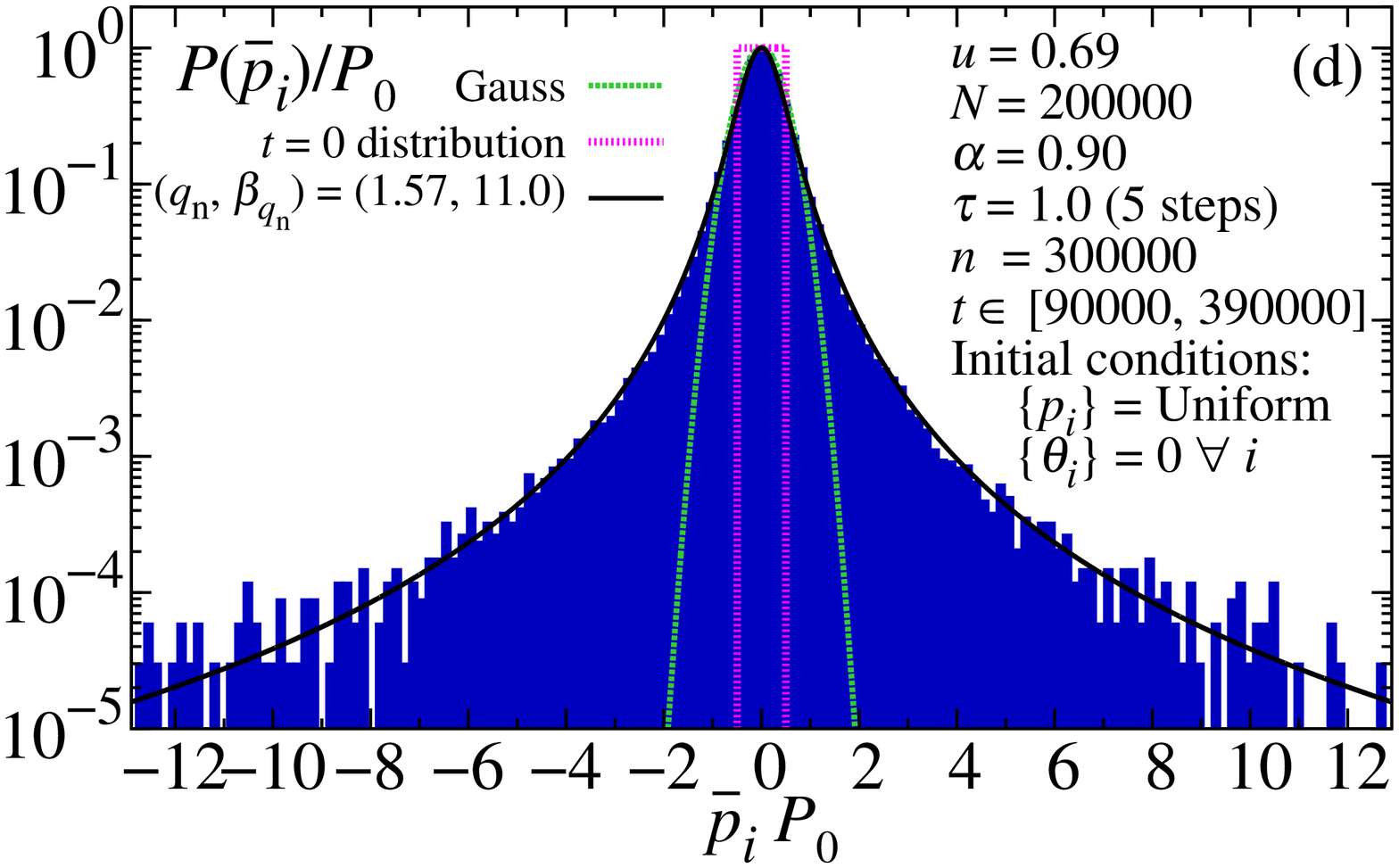}
\end{center}
\caption
{
{\it Top figure}: Time evolution of the kinetic temperature $T$ for $(\alpha, u, N) =(0.90, 0.69, 200000)$
averaged over 50 different realizations, \emph{i.e.}, we run the system 50 times with different seeds of the random number generator for the initial uniform velocity distribution.
After that, a simple (arithmetic) average has been calculated.
One of these 50 realizations was also shown in figure~\ref{fig1}. {\it Panels (a), (b) and (c)}: The averages over these 50 realizations of the momenta
distribution (by $\langle p \rangle$ we denote the average over the realizations) at three distinct times, namely at $t=90000$, $t=240000$ and $t=390000$, as indicated in the top figure.
Let us mention that the Gaussian shape of these averaged distributions remains essentially the same from a few realizations on; in other words, 50 realizations is largely above what would be numerically necessary in order to have invariant results.
{\it Panel (d)}: Time average calculated within the interval $t\in[900000, 390000]$ for a {\it single} realization.
Let us stress that the $q$-Gaussian shape becomes increasingly better (especially in the tails) with increasing $N$, as can be seen by comparing the present figure (d) with figure~\ref{fig2} (top).
Although not shown here, all the corresponding panels (a-d) essentially coincide among them for $\alpha$ large enough (say $\alpha=2$), and they provide a Gaussian shape, as expected for an ergodic system.
}
\label{Fig:Times_Vs_Ensemble_Averages}
\end{figure}

Let us focus on here a very interesting issue, namely the discrepancy between ensemble averages and time averages, which reveals the nonergodicity of the present system whenever the interactions are sufficiently long-ranged. This phenomenon occurs, interestingly enough, {\it even after the kinetic temperature has reached its BG value}.
In figure~\ref{Fig:Times_Vs_Ensemble_Averages}, time and ensemble 
averages are compared for~$\alpha=0.9$ and $N=200000$.
The initial conditions are the same used in the previous results, namely the system starts with magnetization~$M=1$,
which is followed by a violent relaxation, bringing it to the QSS.
Though $N$ is considerably large, the QSS duration is relatively short since $\alpha=0.9$.
After this short QSS period, the system exhibits a crossover to the BG kinetic temperature.
As indicated in the top of figure \ref{Fig:Times_Vs_Ensemble_Averages}, photographies of the velocities of all the particles were taken at $t=90000$, $240000$ and $360000$.
In order to approach an {\it ensemble average}, this procedure was repeated for 50 realizations and the histograms of the velocities were calculated
through simple (arithmetic) means (see figure \ref{Fig:Times_Vs_Ensemble_Averages}(a,b,c)).
For one of these realizations, a {\it time average} was calculated within the interval $t\in[90000,360000]$ (see figure \ref{Fig:Times_Vs_Ensemble_Averages}(d)).

The interesting and deep discrepancy that we observe here between time and ensemble averages (i.e., lack of ergodicity) for relatively small values of $\alpha$ is consistent with what was previously obtained~\cite{PluchinoRapisardaTsallis2007} for ~$\alpha=0$ and small values of $N$ ($N=100$ in that case; see figure~4 in~\cite{PluchinoRapisardaTsallis2007}).
Actually, the study reported in~\cite{PluchinoRapisardaTsallis2007} mainly focused on the QSS regime of the HMF system (see also~\cite{Campa_E_Chavanis_EPJB_2013} as well as \cite{Chavanis_Arxiv_2012}).
It was later clarified~\cite{PluchinoRapisardaTsallis2008}
that, for $(\alpha,u)=(0,0.69)$ and $N$ up to $N\approx20000$,
one can distinguish three classes of QSS events (which were referred to as classes 1, 2 and 3)
whenever the system starts with a water-bag initial condition as described in the above section~\ref{Sec:Numerical_Procedure}.
In~\cite{PluchinoRapisardaTsallis2008}, the $q$-Gaussian one-momentum distributions were obtained for the class~1 events.
It was, however, verified that, for increasingly large $N$, many of the QSS events belong to class~3. Let us mention by the way that the only type of events investigated here precisely are those of class~3: typical $\alpha<1$ patterns are included in figure~\ref{fig1}.

\section{Final remarks}
Summarising, it has been observed for at least one decade that, for $0 \le \alpha<1$, the longstanding QSSs of the present model (i.e., the $\alpha$-XY system of rotators) exhibit anomalous distributions (Vlasov-like for some classes of initial conditions, and different, including $q$-Gaussian-shaped, ones for other classes) for the momenta of the rotators, whereas nothing particularly astonishing was expected to occur once the system had done the crossover to the (presumably stationary) state whose kinetic temperature coincides with that analytically obtained within the BG theory.

The present results (obtained from first principles, \textit{i.e.}, using essentially nothing but Newton's law)
neatly show that, if time is large enough so that the kinetic-temperature crossover has occurred (as illustrated in the Inset of figure~\ref{fig2}), the situation is far \emph{more complex}.
Indeed, robust and longstanding $q$-Gaussian distributions are numerically observed under a wide variety of situations.
The fact that the numerical kinetic temperature be the one predicted within the BG theory 
is sometimes thought as a sufficient condition for standard statistical mechanics to be applicable. However,
the lack of ergodicity caused by the long-range nature of the interactions shows that, at the time range we are focusing on~\cite{remark}, the discussion is more subtle.
Indeed, the shape of the momenta distributions can considerably and  longstandingly differ from Gaussians, and it is only when the correlations become negligible (\textit{i.e.}, when $\tau \gg 1$ and/or $\alpha \gg 1$) that the classical Maxwellian distribution (with $\beta_{q_{\textrm{n}}}^{-1}=T$) is  (numerically) recovered. Similar results are also observed for the angle distributions, as illustrated in figure \ref{figangle}. The full discussion of the angle distributions is out of the scope of the present paper and will be discussed in detail elsewhere.

Let us mention at this point that the breakdown of ergodicity which emerges for $\alpha/d \le 1$~\cite{PluchinoRapisardaTsallis2007, PluchinoRapisardaTsallis2008, FigueiredoRochaAmato2008} as indicated in figure~\ref{Fig:Times_Vs_Ensemble_Averages}
is neatly worthy of further consideration (see also, for a quantum system, \cite{zeilinger}).
Indeed, within the BG framework, not only the kinetic temperature (or the magnetization) must coincide with the canonical prediction,
but a Maxwellian form also is expected in the velocity distribution, independently on whether it is time or ensemble averages  which are being calculated. However, we have shown that long-range interactions make the physical scenario much more delicate.
It is the aim of nonextensive statistical mechanics~\cite{Tsallis2009,Tsallis1988} to provide a 
possible frame for discussing such difficult cases.
Within this generalized theory, a plausible thermostatistical scenario could be as follows.
The stationary state is expected to yield a probability distribution $e_q^{-\beta_q \bar{\cal H}} / Z_q(\beta_q)$ with $Z_q(\beta_q) \equiv \int \rd p_1...\rd p_N \rd\theta_1...\rd\theta_N e_q^{-\beta_q \bar{\cal H}}$.
The index $q$ is expected to characterize universality classes, possibly a function $q(\alpha/d)$ to be different from 1 for $0 \le \alpha/d <1$, and equal to 1 for $\alpha/d \ge 1$.
If this is so, an interesting quantity would of course be the one-momentum marginal probability $P(p_1)= \int \rd p_2...\rd p_N \rd\theta_1...\rd\theta_N e_q^{-\beta_q \bar{\cal H}}/Z_q$. The functional form of $P(p_1)$ is unknown.
A possibility could however be that, in the $N\to\infty$ limit, we simply have $P(p_1) \propto e_{q_{\textrm{m}}}^{-\beta_{q_{\textrm{m}}} p_1^2/2}$, \textit{i.e.}, a $q_{\textrm{m}}$-Gaussian form, where $\textrm{m}$~stands for {\it momentum}.
Indeed, $q$-Gaussians emerge extremely frequently in complex systems
(see, \textit{e.g.},~\cite{Andradeetal2010, RibeiroNobreCurado2012, Ribeiro_Nobre_Curado_EPJB_2012, Casas_Nobre_Curado_PRE_2012, LeoLeoTempesta2010, Tirnakli_et_al_2009, Pluchino_Rapisarda_Tsallis_PRE_2013, Lutz2003, LiuGoree2008, DeVoe2009, MiritelloPluchinoRapisarda2009, DouglasBergaminiRenzoni_PRL_2006};
see also~\cite{cosmicrays, Wong_E_Wilk_Paper_Decadas_2012, KhachatryanCMSCollaboration_1, Aamodt_et_al_ALICECollaboration_1, Abelev_et_al_ALICECollaboration, Aad_et_a_ATLASCollaboration2011, Adare_et_al_PHENIXCollaboration_1} for high energy physics).
The index $q_{\textrm{m}}$ could depend not only on $\alpha/d$, but also, in principle, on $u$.
Similar considerations are in order for the one-angle marginal probability $\Theta(\theta_1)= \int \rd p_1...\rd p_N \rd\theta_2...\rd\theta_N e_q^{-\beta_q \bar{\cal H}}/Z_q$, whose precise analytical form also is unknown.

The present example, with its neat and sensible drift from BG behaviour for short-range interactions to non-BG behaviour for long-range interactions, constitutes a novel illustration of the great thermostatistical richness that a breakdown of ergodicity can cause.
It also serves as an invitation for deeper analysis of the thermal statistics of all those very many models in the literature that are definitively nonergodic (\textit{e.g.}, glasses, spin-glasses, among others), and for which, nevertheless, the BG theory is straightforwardly used without further justification. It illustrates Gibbs' remark \cite{Gibbs1902} that standard statistical mechanics are not justified whenever the canonical partition function diverges (which is the case for long-range interactions, \textit{i.e.}, for $0 \le \alpha \le 1$).

\section*{Acknowledgments}
We acknowledge useful conversations with M. Jauregui, L.G. Moyano, F.D. Nobre, A. Pluchino, A. Rapisarda and L.A. Rios.
We have benefited from partial financial support by CNPq, Faperj and Capes (Brazilian Agencies).

\begin{thebibliography}{99}
\bibitem{fftw}
To implement the Fast Fourier Transform for the $\alpha$-XY model we have used the FFTW library.
\end{thebibliography}

\end{document}